\documentclass[prl,superscriptaddress,twocolumn,showpacs]{revtex4}
\usepackage{amssymb,amsmath}
\usepackage{color}
\usepackage{dsfont}
\usepackage{hyperref}
\usepackage{graphicx}

\definecolor{darkblue}{rgb}{0,0,0.5}
\definecolor{darkred}{rgb}{0.5,0,0}

\hypersetup{
  pdfborder={0,0,0},
  colorlinks=true,
  linkcolor=darkblue,
  citecolor=blue
}

\begin{document}

\title{Ramsey-Bord\'e interferometer for electrons}

\author{Karl-Peter Marzlin}
\affiliation{Department of Physics, St. Francis Xavier University,
  Antigonish, Nova Scotia, B2G 2W5, Canada}
\affiliation{Institute for Quantum Information Science,
        University of Calgary, Calgary, Alberta T2N 1N4, Canada}

\bigskip

\begin{abstract}
A scheme to realize an electron interferometer using 
low-intensity, bi-chromatic laser pulses as beam splitter is proposed.
The splitting process is based on a modification of the Kapitza-Dirac
effect, which produces a momentum kick for electrons with a
specific initial momentum. A full interferometric setup 
in Ramsey-Bord\'e configuration is theoretically analyzed.
\end{abstract}



\maketitle

{\em Introduction.---}
Interferometry with matter has a long history that started
with experiments on electrons \cite{PhysRev.30.705}
shortly after de Broglie hypothesized the existence of matter waves.
Much later
the interference between matter waves was also demonstrated
for neutrons \cite{Rauch1974369}, atoms
\cite{PhysRevLett.66.2689,PhysRevLett.66.2693},
molecules \cite{PhysRevLett.88.100404}
and ions \cite{0034-4885-73-1-016101}. In most of these
experiments the particle beam is split by mechanical means such as gratings
or crystal lattices. In atom interferometry, however, 
one can also use laser light to split the beam
\cite{PhysRevLett.67.177,PhysRevLett.67.181} by transferring
the momentum of absorbed photons to the atoms.
This technique enables temporal control of the splitting process,
makes the set-up of an interferometer very flexible, and has
had an enormous influence on atom optics, metrology, and
quantum information.

Atom interferometers are excellently suited for many purposes,
but some phenomena, such as the Aharonov-Bohm effect \cite{aharonov59},
require the use of charged particles. It would be desirable to use fields as
beam splitters for electrons and thus acquire the same flexibility
in setting up an interferometer. 
For neutral particles in magnetic fields the Stern-Gerlach effect may
be used, but for electrons the Lorentz force makes it very
difficult to design a beam splitter
\cite{1367-2630-13-6-065018,PhysRevLett.86.4508}.
For electrons in electric fields, the Kapitza-Dirac effect \cite{PSP:1736920,RevModPhys.79.929} 
may be suitable, which has been observed experimentally in the 
high light intensity \cite{PhysRevLett.61.1182} and low light
intensity regime \cite{Nature413-142}. Current efforts to design 
temporal lenses for electrons require very high light intensities
\cite{Baum20112007,Hilbert30062009}.

In this paper, I propose to realize beam splitters for electrons
by using two counter-propagating laser pulses
with a specific frequency difference.  
This modification of the Kapitza-Dirac effect is related to
diffraction without grating \cite{PhysRevLett.92.223601}
but uses a resonance condition to achieve a specific momentum
transfer from the light field to the electrons. I will then show how
such beam splitters may be used to set up a complete
interferometer of Ramsey-Bord\'e 
type \cite{Borde198910,PhysRevLett.67.177,PRA47:4441,PRA50:2080}.

{\em Kapitza-Dirac beam splitter.---}
The principle behind the current proposal is conservation of energy and momentum
in the interaction between an electron and two counter-propagating
laser fields of frequency $\omega_i$ ($i=1,2$) and wavenumber
$k_i=\omega_i/c$. If an electron absorbs a photon from laser 1 and
subsequently emits a photon stimulated by laser 2, its momentum
will change by $2\hbar k_L$
and its energy by $-\hbar \Delta \omega$, where
$\Delta \omega\equiv\omega_2-\omega_1$
and $k_L\equiv \left(k_1+k_2\right)/2$. If for a given
initial momentum the change in kinetic energy is equal
to $\hbar \Delta \omega$, this process will be resonant
and can be accomplished with light intensities as used
for Bragg scattering \cite{PhysRevLett.89.283602}.

To describe this effect we consider the Hamiltonian
  $ \hat{H} =  ( -i\hbar \nabla -q \vec{ A} )^2/(2m)$
with vector potential 
$\vec{ A}=\vec{ \epsilon}( A^{(+)}+A^{(-)})$, where
\begin{align} 
  A^{(+)}&= -\frac{i E_1 e^{i k_1 z-i t \omega _1+i\theta_1}}{4 \omega _1}-\frac{i E_2 e^{-i
   k_2 z-i t \omega _2+i\theta_2}}{4 \omega _2}
\label{eq:KDvectorpot}\end{align} 
is the positive-frequency part of the field and $ A^{(-)}= ( A^{(+)})^*$.
$E_i$ is the electric field amplitude of laser $i$.
The unit vector $\vec{ \epsilon}$ describes the polarization
direction in the $x$-$y$-plane and $\theta_i$ are phase factors.
By expanding the Hamiltonian we obtain terms that are linear or
quadratic in the vector potential.
The optical frequencies $\omega_i$ are much larger than any other frequency
scale in our system, so that linear terms and
terms of the form $(A^{(\pm)})^2$
 are rapidly oscillating. After time averaging
\cite{Fedorov1974205}, only terms of the form
 $A^{(+)} A^{(-)}$ survive and we obtain the averaged Hamiltonian
\begin{align} 
  \hat{H}_\text{avg} &= -\frac{ \hbar^2}{2m} \Delta +
  \frac{ q^2}{8m} \Big ( 
  \frac{E_1^2}{2 \omega _1^2}+\frac{E_2^2}{2 \omega _2^2}
\nonumber \\ &\hspace{5mm} +
  \frac{E_2 E_1
   \cos \left(2 k_L z+\Delta \omega  t-\Delta\theta\right)}{ \omega
   _1 \omega _2}
  \Big ),
\end{align} 
with $\Delta\theta=\theta_2-\theta_1$. Restricting
our considerations to the $z$-direction and performing
a spatial Fourier transformation yields 
\begin{align} 
  i \partial_t \psi(t,k)&=
   \left(\frac{ \hbar k^2}{2m}
   + g_1^2 +g_2^2\right)
  \psi (t,k) 
  +  g_1 g_2  \Big ( e^{i \Delta \omega  t-i\Delta\theta} 
\nonumber \\ &\hspace{0mm}\times
    \psi (t,k-2 k_L)
   + e^{-i  \Delta \omega  t+i\Delta\theta}\psi(t,k+2 k_L) \Big ) ,
\end{align} 
with $g_i \equiv q E_i/(4\omega_i \sqrt{m\hbar})$.
The coupling between discrete momentum components
in this equation is similar to the problem of calculating
band gaps for particles in periodic potentials. Following
standard methods we introduce the quasi
wavenumber $\bar{k}\in [-k_L,k_L]$ and express the
electron wavenumbers as $k = \bar{k}+ 2n k_L$ for
$n\in \mathds{Z}$. Setting
$\psi_n(t,\bar{k}) \equiv  \psi (t,\bar{k}+2n k_L)$ we then obtain
\begin{align} 
  i \partial_t \psi_n(t,\bar{k})&=
   \left(\frac{ \hbar (\bar{k}+2nk_L)^2}{2m}
   + g_1^2 +g_2^2\right)
  \psi_n(t,\bar{k}) +  g_1 g_2 
\nonumber \\ &\hspace{-10mm}
  \times\Big ( e^{i \Delta \omega  t-i\Delta\theta} 
    \psi_{n-1} (t,\bar{k})
   + e^{-i  \Delta \omega  t+i\Delta\theta}\psi_{n+1}(t,\bar{k}) \Big ) .
\end{align} 
A unitary transformation
\begin{align} 
    \psi_n(t,\bar{k}) &=
   e^{-it \left(n \Delta \omega + g_1^2+g_2^2+ \frac{ \hbar}{2m} \bar{k}^2
     \right)+i n \Delta\theta}
  \bar{\psi }_n(t,\bar{k})
\label{eq:unitransf}\end{align} 
yields 
\begin{align} 
  i \partial_t \bar{\psi }_n(t,\bar{k}) &=
   n \left(  
      n  \frac{2 \hbar  k_L^2}{m}+
    \frac{2 \hbar \bar{k}  k_L}{m}
    +\Delta \omega \right) \bar{\psi}_n(t,\bar{k}) 
\nonumber \\ &\hspace{5mm}
+ g_1 g_2 \left ( \bar{\psi }_{n-1}(t,\bar{k})+ \bar{\psi }_{n+1}(t,\bar{k}) \right ).
\label{eq:bichromPulsesDgl}\end{align} 

For $\Delta \omega=0$ this equation describes 
Bragg scattering \cite{PhysRevLett.89.283602}, but we
can use $\Delta \omega$ to make the interaction resonant
for a specific initial momentum. For concreteness we consider
an initial electron wavepacket with mean momentum zero
and spatial width $w$, 
\begin{align} 
  \bar{\psi}_0(0,\bar{k}) &=
  e^{-\bar{k}^2 w^2} 
   2^{\frac{ 1}{4}} w^{\frac{ 1}{2}}
  \pi^{-\frac{ 1}{4}} ,
\label{eq:momentumComponents}\end{align} 
such that $w k_L \ll 1$, and 
$\bar{\psi }_n(0,\bar{k}) = 0$ for $n\neq 0$. The energy difference
between the state $n=0$ and the states $n=\pm 1$ is then given by
\begin{align} 
  \Delta E_{\pm 1,0} &= 
      \omega_\text{rec} \pm 
    \frac{2 \hbar \bar{k}  k_L}{m}
    \pm \Delta \omega 
\; \approx   \omega_\text{rec}
    \pm \Delta \omega ,
\end{align} 
with
$   \omega_\text{rec} = 2 \hbar  k_L^2/m$
the recoil shift.
For $\Delta \omega = \mp \omega_\text{rec}$
the $n=0$ state is resonant with the state $n=1$ or $n=-1$, respectively.
In the limit of weak coupling, $g_1 g_2 \ll  \omega_\text{rec}$,
all other components $\bar{\psi }_n(t,\bar{k}) $ can be neglected, so
that, e.g., for $\Delta\omega=-\omega_\text{rec}$
the Schr\"odinger equation reduces to
\begin{align} 
  i \partial _t 
 \left ( \begin{array}{c}
          \bar{\psi }_0  (t)  \\ \bar{\psi }_1 (t)
           \end{array} \right ) &=
  \left ( \begin{array}{cc}
           0  & g_1 g_2 \\  g_1 g_2 &  \frac{ 2}{m} \hbar k_L \bar{k}
           \end{array} \right ) 
  \left ( \begin{array}{c}
          \bar{\psi }_0 (t)   \\ \bar{\psi }_1 (t)
           \end{array} \right ).
\end{align} 
For $\bar{k}=0$ and $t=\pi/(4 g_1 g_2)$ the solution is given by
\begin{align} 
 \left ( \begin{array}{c}
          \bar{\psi }_0  (t)  \\ \bar{\psi }_1 (t)
           \end{array} \right ) &= \frac{ 1}{\sqrt{2}}
  \left ( \begin{array}{cc}
           1  & -i \\  -i &  1
           \end{array} \right ) 
  \left ( \begin{array}{c}
          \bar{\psi }_0 (0)   \\ \bar{\psi }_1 (0)
           \end{array} \right ),
\label{eq:KDbsResult}\end{align} 
which corresponds to a perfectly balanced beam splitter.
For $\bar{k}\neq 0$ the beam splitter will not be perfectly balanced,
but if $g_1 g_2 \gg 2\hbar k_L\, \Delta \bar{k}/m$, where
$\Delta \bar{k} =1/(2w)$ is the width of the wavepacket in momentum space,
Eq.~(\ref{eq:KDbsResult}) still provides an excellent approximation.

To estimate the feasibility of this proposal
we consider a laser wavelength of 1064 nm, so that
$\omega_\text{rec}= 2\pi \times 1.3$ GHz. The generation of two 
phase-locked laser pulses with such a detuning is well within
the range of current experimental techniques \cite{0957-0233-20-5-055302}.
For lasers of equal intensity, $g_1=g_2$, the weak 
coupling condition then results in the constraint
$
  q^2 I/(32 \omega^2 m \hbar c \varepsilon_0) \ll \omega_\text{rec},
$ 
where $I=2c\varepsilon_0 E^2$ denotes the light intensity. 
In our case this implies $I\ll 8 \text{W}/\mu\text{m}^2$, 
which is considerably less than the
intensity used in Ref.~\cite{Nature413-142}. For an intensity
of $I=0.5 \text{W}/\mu\text{m}^2$, i.e., $g_1g_2=2\pi\times 80$ MHz,
the pulse duration is $\pi/(4 g_1g_2) \approx 1.5$~ns. 
The spatial width of the
electron wavepacket needed to obtain a balanced beam splitter is then
$w \gg 1 \mu$m. 

\begin{figure}
\begin{center}
\includegraphics[width=7cm]{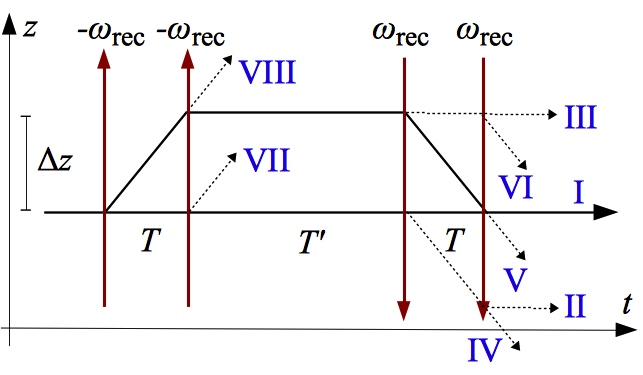}
\caption{\label{fig:RamseyInterferometer}
Setup of a Ramsey-Bord\'e interferometer for electrons. A sequence
of four bichromatic laser pulses (red vertical lines), each of which
is composed
of two counter-propagating waves detuned by a multiple of the recoil shift
$\omega_\text{rec}$,
 is used to split and recombine
the electron beam. The beam is split into eight partial beams, which are
represented by arrows with a roman number.
Solid black lines correspond to a spatially closed interferometer geometry.}
\end{center}
\end{figure}

{\em Ramsey-Bord\'e interferometer.---}
Kapitza-Dirac beam splitters could be utilized
to implement a Ramsey-Bord\'e interferometer
for electrons, with a setup as shown
in Fig.~\ref{fig:RamseyInterferometer}.
The electron beam is split and recombined by four bichromatic laser
pulses. The first two pulses are detuned by $\Delta\omega = -\omega_\text{rec}$,
thus coupling electrons with an initial momentum $p\approx 0$ to
$p+2\hbar k_L$. During the time $T$ between these pulses
the electron beam is spatially split into two beams, which after the
second pulse are separated by a distance
$  \Delta z = T\Delta v = 2\hbar k_L T/m$.
The third and fourth pulse are detuned by $\Delta\omega =
\omega_\text{rec}$,  so that 
$p\approx 0$ is coupled to $p-2\hbar k_L$. This enables us to
recombine (parts of) the electron beam, leading to a closed
interferometer that corresponds to output I (and V) in Fig.~\ref{fig:RamseyInterferometer}.

To understand the details of this interferometer we
describe the electrons by the time-dependent Schr\"odinger equation.
For simplicity we will only consider the motion along the
$z$-axis (perpendicular to the initial beam direction).
During the free evolution for a time $T$ between two pulses the
dynamics is governed by Hamiltonian $\hat{H}_0 = - \hbar^2\partial_z^2 /(2m)$.
During the time $T'$ between the beam splitting and recombination
processes we admit a constant electric field $E$ to study its effect
on the phase difference between the beams. The corresponding
Hamiltonian is $\hat{H}_a = \hat{H}_0  -m a z $, with acceleration
$a=e E/m$.
Inhomogeneous electric fields could be studied with the methods of
Ref.~\cite{PRA47:4441}.

To describe the interference experiment, we start with an initial
state of the form
$
   \psi_\text{init}(z)=\int dp \,
   \tilde{\psi}(p) 
   \phi_{p}(z) ,
$
with momentum eigenstates
 $ \phi_{p}(z)= e^{i p z/\hbar} /\sqrt{2\pi\hbar}$
and $\int dp\, | \tilde{\psi}(p)|^2=1$. 
The evolution between the laser pulses can then be described by
the unitary transformations $\exp(-i T \hat{H}_0/\hbar) \phi_{p}(z)
= \exp(-i T E(p)/\hbar) \phi_{p}(z)$, where $E(p) = p^2/(2m)$, and
\begin{align} 
  e^{-\frac{ i}{\hbar} T' \hat{H}_a} \phi_{p}(z) &= e^{-i \tau(p)} \;
  \phi_{p+m a T'}(z),
\label{eq:schroedSolAccel}
\\
   \tau(p) &= \frac{ 1}{\hbar} \int_0^{T'} dt'\, E(p+m a t').
\label{eq:tauDef}\end{align} 
A simplified description of Eq.~(\ref{eq:KDbsResult})
for the first two pulses, with $\Delta \omega=-\omega_\text{rec}$, 
can be given by the unitary transformation
\begin{align} 
    \hat{U}_{-} \phi_{p}(z)&=
   \frac{ 1}{\sqrt{2}} \left (
  \phi_{p}(z) + \phi_{p+2\hbar k_L}(z)
  \right )
\\ 
     \hat{U}_{-} \phi_{p+2\hbar k_L}(z)&=
   \frac{ 1}{\sqrt{2}} \left (
  -\phi_{p}(z) + \phi_{p-2\hbar k_L}(z)
  \right ).
\label{eq:UtrafoSimpl}\end{align} 
For the second pair of pulses we chose $\Delta \omega=+\omega_\text{rec}$,
which results in a similar unitary transformation
$\hat{U}_{+}  $, which is equal to $\hat{U}_-$ with $k_L$ replaced by $-k_L$. 
These expressions are only 
valid for momenta $|p| \ll \hbar k_L$.
Eq.~(\ref{eq:UtrafoSimpl}) is not accurate but captures
the essential physics of each pulse. A more complete treatment
will be given below.

The final state of the electrons after passing through the
interferometer can be found by concatenating all unitary
transformations.
This splitting process, 
corresponding to the pair of pulses labeled by $- \omega_\text{rec}$ in 
Fig.~\ref{fig:RamseyInterferometer}, can be
described by a unitary operator $\hat{U}_\text{split} =\hat{U}_{-}   
\exp(-i T \hat{H}_0/\hbar)\hat{U}_{-}  $. Analogously, the recombination
process can be described through the unitary operator
$\hat{U}_\text{rcmb} =\hat{U}_{+}   
\exp(-i T \hat{H}_0/\hbar) \hat{U}_{+}  $, which results in
the final state  
\begin{align} 
  \hat{U}_\text{rcmb}&e^{-\frac{ i}{\hbar} T' \hat{H}_a}\hat{U}_\text{split}  \psi_\text{init}(z) = 
  \frac{ 1}{4} \int dp \,  \tilde{\psi}(p)     \phi_{p+maT'}(z) 
\nonumber \\ & \hspace{0cm}\times 
   e^{-i \tau(p,T')}   \Big (
  e^{-\frac{ i T}{\hbar} E (p+2\hbar k_L) }  
  e^{-\frac{ i T}{\hbar} E(p-2\hbar  k_L+m a T') }
\nonumber \\ &
   + e^{-\frac{ i T}{\hbar} E(p) }  
  e^{-\frac{ i \tilde{T}}{\hbar} E(p+m a T') }
\Big ) + \text{rest}.
\label{eq:finalStateNonrel}\end{align} 
In this expression, ``rest'' refers to seven terms that are similar
to the ones displayed and correspond to
partial beams represented by dashed arrows in 
Fig.~\ref{fig:RamseyInterferometer}. 
The two terms that are displayed
correspond to the solid black lines, which realize a
spatially closed interferometer geometry. In an atomic
Ramsey-Bord\'e interferometer, this geometry is useful
because the phase difference between the two beams
is not influenced by the Doppler effect \cite{PRA53:312}.
This is also the case for electrons, for which the phase difference
is given by
\begin{align} 
  \Delta \varphi &=
   \frac{ 4}{m}\hbar k_L^2 T' -2 a k_L T T'.
\label{eq:nrPhaseShift}\end{align} 
This differs from previous results (e.g., Eq.~(45) of
Ref.~\cite{PRA50:2080}) because the momentum transfer $2\hbar k_L$
is twice as big as in atom interferometers and
because we assumed that the acceleration is only present during 
$T'$.

With the simplified description (\ref{eq:UtrafoSimpl})
of Kapitza-Dirac beam splitters and initial state
(\ref{eq:momentumComponents}) one can evaluate the final state
Eq.~(\ref{eq:finalStateNonrel}) in a closed form. 
The full solution is rather involved and will not
be displayed here. However, it can be shown that the final spatial 
wavefunction of partial beam I corresponds to a (dispersing) 
Gaussian wavepacket 
that is centered at $z= \frac{ 1}{2} a T'^2+ a T T'$. As is to be
expected because of Ehrenfest's theorem, its mean
position follows a classical trajectory that is determined by
momentum kicks due to the laser pulses combined with
free or accelerated motion between the pulses. All
other partial beams do also correspond to Gaussian
wavepackets that follow classical trajectories.
\begin{figure}
\begin{center}
\includegraphics[width=7.5cm]{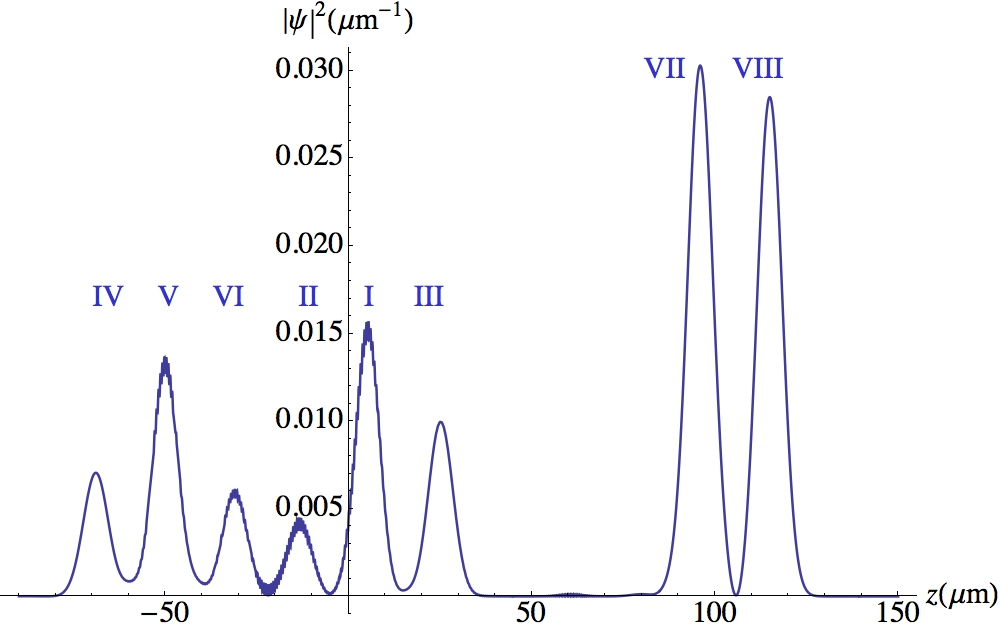}
\caption{\label{fig:RamseyNonrel}
Spatial interference pattern for non-relativistic electrons. The
roman numbers of each wavepacket corresponds to the
partial beams depicted in Fig.~\ref{fig:RamseyInterferometer}. }
\end{center}
\end{figure}

To provide a more accurate description of the beam splitting process
we also solved the Schr\"odinger equation numerically, including the
Kapitza-Dirac effect as described by Eq.~(\ref{eq:bichromPulsesDgl}).
The spatial profile of the electron probability density after passing
through the interferometer is shown in Fig.~\ref{fig:RamseyNonrel},
where we have used an initial wavepacket of the form
(\ref{eq:momentumComponents}) with a mean width of $w=3\mu$m, a
laser wavenumber of $k_L= 2\pi/1064$~nm and an acceleration of 
$a= 10^{10} \text{m} /\text{s}^{2}$. The time $T$ between the two
pulses of the beam splitter and recombiner is $T=12$~ns and the
electrons are accelerated for $T'=10$~ns so that their final velocity
transverse to the direction of the beam is 
100~m/s \footnote{For much larger transverse velocities the Doppler
effect will lead to a detuning between the two counter-propagating
laser fields, which would modify the resonance condition.}. 
To resolve all partial beams
(for clarity of presentation)
we have added another free evolution for a time $T''=40$ ns after the 
last laser pulse. 

The two partial beams that form a closed geometry
produce peak I in Fig.~\ref{fig:RamseyNonrel}. The offset of this peak
from the origin corresponds to the distance travelled due to the acceleration. 
The position of all
other peaks is mainly determined by the sequence of momentum
transfers that they obtain. For instance, peak VIII obtains a momentum
of $2\hbar k_L$ at the first bichromatic pulse. Because all other
partial beams either obtain this momentum transfer later, or obtain no
or negative momentum transfer, it is pulse VIII that will travel
to the rightmost position. If we had set $T''=0$, i.e., if the
interference pattern was observed right after the last laser pulse,
wavepackets IV, V, and VI would be overlapping with wavepackets II, I, and
III, respectively. These partial beams only differ by a momentum kick
generated by the last laser pulse.

The position of the partial beams in
Fig.~\ref{fig:RamseyNonrel} is not in perfect agreement
with the classical trajectories discussed above. The reason is that
the latter ignores the finite duration of the interaction between
the laser pulses and the electrons. If we assume that during the
interaction time each partial beam moves with the average
of it momentum before and after the pulse, the agreement
between analytical and numerical solution is excellent.

Ramsey-Bord\'e interferometers for atoms and electrons
have a similar geometry but differ significantly in some details.
First, the atom interferometer has more partial
beams than the electron interferometer and even includes a
second pair of beams with a closed geometry (see Fig.~1 of
Ref.~\cite{PRA50:2080}, for instance). The reason is that in
an atom interferometer the resonance condition is determined
by atomic energy levels. Hence, light transfers
momentum to an atom regardless of the value of its
center-of-mass momentum. On the other hand, 
in a Kapitza-Dirac beam splitter 
the kinetic energy determines the resonance condition, so that
only electrons with a specific initial momentum will 
resonantly interact with the bichromatic laser pulses. For this reason, 
partial beams VII and VIII will not be affected by the second pair of
laser pulses, resulting in a reduced number of partial beams.

Second, it is worth to remark that there are no spatial interference
fringes in Fig.~\ref{fig:RamseyNonrel} because for
our choice of $T,T',T''$ all partial beams are separated. 
However, the phase difference (\ref{eq:nrPhaseShift}) has a strong
influence on the relative intensities of the partial beams.
This situation is similar in atom interferometers, where
the probability for the atoms to be in the excited state instead
of a spatial interference pattern is observed \cite{PhysRevLett.67.177}.
This probability is a function of the partial beam intensities and
thus is sensitive to a phase difference.

{\em Conclusion.---}
In this paper I have proposed to use a modification
of the Kapitza-Dirac effect to devise beam splitters
and a Ramsey-Bord\'e interferometer for electrons.
Numerical simulations suggest that such an experiment
may be realized using current technology. 
Field-based electron beam splitters would allow 
for a much more flexible setup that could lead to 
novel applications of electron interferometers.
For instance, ``figure 8'' interferometer geometries
could be used to test the phase shift induced
by a spatial variation of an electric field, rather than
its amplitude \cite{PRA53:312,physics:0604082}.
Many more applications that involve electric and magnetic fields
are possible and will be explored in future publications.

\acknowledgments
This project was funded by NSERC, AceNet and  
a UCR grant from St.~Francis Xavier University.

\bibliographystyle{apsrev4-1}
\bibliography{/Users/pmarzlin/Documents/literatur/kpmJabRef.bib}
\end{document}